\documentclass[useAMS,usenatbib]{mn2e}
\usepackage{latexsym,graphicx}
\usepackage{amsmath}

\newcommand\msun{\, \rm M_\odot}
\newcommand\pc{{\, \rm pc}}
\newcommand\spc{{\, \rm pc^{-2}}}
\newcommand\cpc{{\, \rm pc^{-3}}}
\newcommand\yr{{\, \rm yr}}
\newcommand\myr{{\, \rm Myr}}
\newcommand\tpr{{\, T_{\rm NP}}}
\newcommand\tecc{{\, T_{\rm ecc}}}
\newcommand\mcusp{{\, M_{\rm cusp}}}
\newcommand\mring{{\, M_{\rm ring}}}

\def\gtsima{$\; \buildrel > \over \sim \;$}
\def\ltsima{$\; \buildrel < \over \sim \;$}
\def\gtrsim{\lower.5ex\hbox{\gtsima}}
\def\lesssim{\lower.5ex\hbox{\ltsima}}

\title[Eccentric disc instability in the Galactic Centre]
{Eccentric disc instability in stellar discs formed from inspiralling
  gas clouds in the Galactic Centre}

\author[Gualandris et al.]  
  {Alessia Gualandris$^{1,2}$\thanks{E-mail:
      alessiag@mpa-garching.mpg.de}, Michela Mapelli$^{3}$ and Hagai B.~Perets$^{4,5}$\\
  $^{1}$Max-Planck Institut f\"{u}r Astrophysik, Karl-Schwarzschild-Str. 1, D--85741 Garching, Germany\\
  $^{2}$Department of Physics and Astronomy, University of Leicester, Leicester, LE1 7RH, United Kingdom\\
  $^{3}$INAF-Osservatorio Astronomico di Padova, Vicolo dell'Osservatorio 5, I--35122, Padova, Italy\\
  $^{4}$Harvard-Smithsonian Center for Astrophysics, 60 Garden St.,
  Cambridge, MA, USA\\
  $^{5}$Technion - Israel Institute of Science, Haifa, Israel 32000\\
   }

\begin{document}

\date{}

\maketitle

\begin{abstract}
The inspiral of a turbulent molecular cloud in the Galactic centre may
result in the formation of a small, dense and moderately eccentric gas
disc around the supermassive black hole (SMBH). Such a disc is
unstable to fragmentation and may lead to the formation of young
massive stars in the central parsec of the Galaxy. Here we perform
high-accuracy direct summation $N$-body simulations of a ring of
massive stars (with initial semi-major axes $0.1\le{}a/{\rm
  pc}\le{}0.4$ and eccentricities $0.2\le{}e\le{}0.4$), subject to the
potential of the SMBH, a stellar cusp, and the parent gas disc, to
study how the orbital elements of the ring evolve in time.  The
initial conditions for the stellar ring are drawn from the results of
previous simulations of molecular cloud infall and disruption in the
SMBH potential.  While semi-major axes do not evolve significantly,
the distribution of eccentricities spreads out very fast ($\approx
1\myr$) as a consequence of cusp precession. In particular, stellar
orbits with initial eccentricity $e>0.3$ ($e<0.3$) tend to become even
more (less) eccentric, resulting in a bimodal eccentricity
distribution. The distribution is qualitatively consistent with that
of the massive stars observed in the Galactic centre's clockwise disc.
\end{abstract}

\begin{keywords}
black hole physics -- methods: N-body simulations -- stars: kinematics and
dynamics -- Galaxy: centre
\end{keywords}

\section{Introduction}

Despite the strong tidal field exerted by the supermassive black hole
(SMBH) in the Galactic centre (GC, \citealt{morris1993}), more than a
hundred young massive stars are observed in the central parsec of the
Galaxy. 

The majority are O-type and Wolf-Rayet (WR) stars, of which
about two thirds show clockwise motion when projected on the plane of the sky
while the remainder show counterclockwise motion
\citep{paumard2006,nay2006,martins2007,lu2009,bartko2009}. The WR/O
stars in the larger group orbit the SMBH in a thin disc, 
usually referred to as the clockwise (CW) disc, with an
estimated age of $\sim 6\myr$ and an average eccentricity of $\sim
0.4$. The smaller group of stars on counter-rotating orbits may indicate
the presence of an additional dissolving disc
\citep{bartko2009,lb2009}. The CW disc shows a sharp inner truncation
at about one arcsecond (or $0.04\pc$).

The S-cluster \citep{scho2003,ghez2003,ghez2005,eisen2005,
  gill2009} located in the innermost arcsecond is a group of about 20
B-type stars moving on eccentric and randomly oriented orbits, with
estimated ages of $20-100\myr$.  The origin of the S-cluster is
largely controversial, with main models including the inspiral of a
young star cluster \citep{gerhard2001,km2003,
  mcmillan2003,gr2005,levin2005,fujii2008}, possibly hosting an
intermediate-mass black hole
\citep{hm2003,bh2006,mgm2009,gm2009,ggm2010,fujii2009} and the tidal
disruption of binaries on low-angular momentum orbits
\citep{gq2003,perets2007, perets2009, pgkma2009, antonini2010,
  pg2010}.

 The formation of the CW disc, which represents the target of this
 work, is also debated, but is often attributed to in-situ star
 formation from an infalling gas cloud \citep{morris1993, nc2005,
   br2008, mapelli2008, wardlezadeh2008, hobbsnayakshin2009, alig2011,
   mapelli2012}.  In particular, \citet{mapelli2012}, hereafter M12,
 perform $N$-body/Smooth Particle Hydrodynamics (SPH) simulations of
 the infall of a turbulent molecular cloud toward the GC and study the
 formation of stars in the region. They find that the tidal disruption
 of the cloud results in the formation of a small (radius $<0.5\pc$),
 dense ($>10^8$ atoms cm$^{-3}$) and eccentric ($e\sim 0-0.6$) gas
 disc around the SMBH. The simulations by M12 are the first following
 the fragmentation of the gas disc into self-bound clumps. Stars
 originate from such clumps in a ring at a distance of $0.1-0.4\pc$
 with moderately eccentric orbits ($e\sim0.2-0.4$).  The mass function
 of the stars is top-heavy if the local background temperature is
 sufficiently high ($\ge{}100$ K) and if the parent cloud is
 sufficiently massive ($\ge{}10^5\msun$). These properties are in good
 agreement with observations of the CW disc
 \citep{paumard2006,bartko2009}, and lend further support to the
 in-situ formation model for the disc.  The simulations by M12,
 however, follow the evolution of the system for only about $0.5\myr$
 since the formation of the gaseous disc, whereas the estimated age of
 the CW disc is of about $6\myr$.

Here we use high-accuracy direct summation $N$-body simulations to
follow the evolution of the stellar ring for $10\myr$ under the effect
of the combined gravitational potential due to the SMBH, the stellar
cusp, and the residual gas disc.  We stress that this is the first
calculation ever performed to follow both the short term star
formation phase and the long term evolution of the disc in a
consistent manner. We find that the disc retains its initial inner and outer
radius, and that stars remain largely confined to
the initial ring plane, which is aligned with the gas
disc. Interestingly, we observe evidence of the eccentric disc
instability described by \citet{madigan2009}, according to which
precession of the orbits due to the presence of the stellar cusp
induces coherent torques that drive orbital eccentricities away from
their initial values. As a result, a bimodal eccentricity distribution
is established which is consistent with observations of the CW disc in
the GC \citep{bartko2009}.

\section{N-body simulations}

\begin{figure}
  \begin{center}
    \includegraphics[height=8cm]{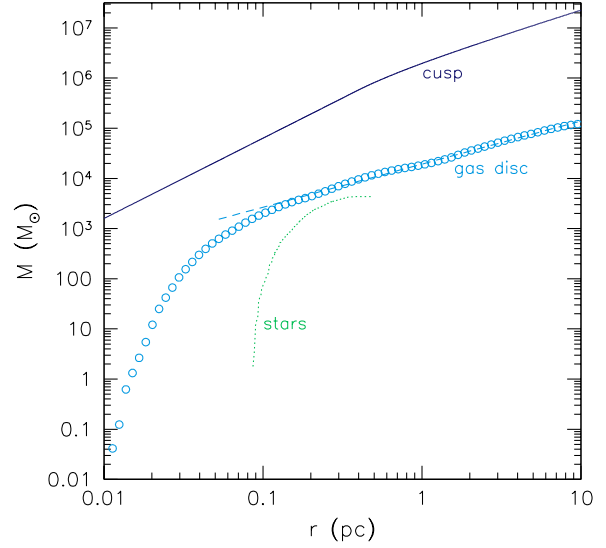}
  \end{center}
  \caption{Solid line: mass profile of the stellar cusp whose
    gravitational effect is included in the simulations in the form of an
    analytic external potential. Points: mass profile of the gaseous
    disc obtained by the SPH simulations of M12. Dashed line: power-law fit
    to the gas disc profile adopted to generate the external potential
    by means of Eq.\,\ref{eq:discpot}. Dotted line: mass profile of
    stellar ring as obtained by M12 and used as initial conditions for
    the integrations.}
  \label{fig:massp}
\end{figure}

We use the direct summation $N$-body code $\phi$GRAPE
\citep{harfst2007} modified to include the effects of an external
potential due to the stellar cusp and the gas disc.
We adopt a mass of $M_{\bullet} = 3.5\times10^6\msun$ for the SMBH and the broken
power-law profile of \citet{genzel2003} for the stellar cusp density
\begin{equation}
\label{eq:cusp}
\rho(r) = \rho_0 (r_0/r)^{\gamma}
\end{equation}
with $r_0 = 10'' \sim 0.39\pc$, $\rho_0 = 1.2\times10^6\msun\cpc$,
$\gamma = 1.4$ for $r < r_0$ and $\gamma = 2.0$ for $r\geq r_0$.
For the residual gas disc we adopt a projected density profile of the
type
\begin{equation}
\label{eq:disc}
\Sigma(R) = \Sigma_0 (R_0/R)^{\alpha}
\end{equation}
with parameters $\alpha = 1.14$, $R_0 = 1\pc$ and $\Sigma_0 =
2.64\times10^3 \msun\spc$ derived from fitting a power-law profile to
the gas remaining after star formation in the simulation of M12.  The
mass profile of the stellar cusp and gas disc are shown in
Fig.\,\ref{fig:massp}.
The gravitational potential generated by a thin disc of known
$\Sigma(R)$ is given by the integral \citep[][Eq.2-142b]{BT1987}
\begin{equation}
\label{eq:discpot}
\Phi(R,z) = - \frac{2G}{\sqrt{R}} \int_0^{\infty} \mathcal{K}(k)\,k\,\Sigma(R^{\prime})\,\sqrt{R^{\prime}}\,dR^{\prime}
\end{equation}
where 
\begin{equation}
k = \sqrt{\frac{4RR^{\prime}}{\left(R+R^{\prime}\right)^2+z^2}}
\end{equation}
and $\mathcal K(k)$ represents the complete elliptic integral of the
first kind.  Differentiating Eq.\,\ref{eq:discpot} with respect to $R$
and $z$ gives
\begin{eqnarray}
\label{eq:dphidr}
\frac{\partial \Phi}{\partial R} (R,z) & = & \frac{G}{R^{3/2}}
\int_0^{\infty} 
\left[ \mathcal{K}(k) - \frac{1}{4}\left(\frac{k^2}{1-k^2}\right) \mathcal{E}(k)
  \right.\nonumber\\
&& \times  \left. \left(\frac{R^{\prime}}{R}-\frac{R}{R^{\prime}} + \frac{z^2}{RR^{\prime}}\right)  \right]
k \, \Sigma(R^{\prime}) \sqrt{R^{\prime}}\,dR^{\prime} 
\end{eqnarray}
and
\begin{equation}
\frac{\partial \Phi}{\partial z} (R,z) = 4Gz \int_0^{\infty}
\frac{\mathcal {E}(k)}{1-k^2} \frac{\Sigma(R^{\prime}) \,R^{\prime}
  \,dR^{\prime} }{\left[\left(R+R^{\prime}\right)^2 + z^2\right]^{3/2}}\,,
\end{equation}
where $\mathcal{E}(k)$ represents the complete elliptic integral of
the second kind.  The equations for the gravitational potential and
forces can be solved numerically on a cylindrical coordinate grid.

\begin{figure}
  \begin{center}
    \includegraphics[height=8cm, angle=-90]{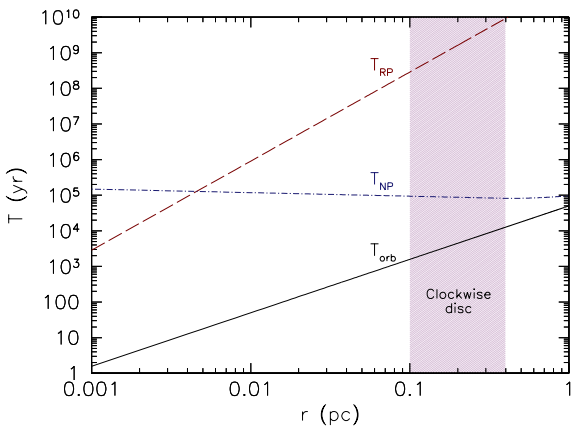}
  \end{center}
  \caption{Relevant timescales for the ring stars in the GC: orbital
    period (solid line), timescale of GR precession (dashed lines) and
    timescale for Newtonian precession (dot-dashed line).  The shaded
    area indicates the approximate location of the CW disc. An orbital
    eccentricity of $0.3$ is assumed.}
  \label{fig:times}
\end{figure}

Relevant timescales for the stars are shown in Fig.~\ref{fig:times} as
a function of distance from the SMBH.
 At distances of $0.1-0.4\pc$, precession due to general relativity
 (GR) is not important on timescales of $10\myr$, and we
 therefore neglect it in the simulations. On the other hand, Newtonian
 precession due to the stellar mass distributed within the orbits of
 the stars has a timescale of \citep[e.g.][]{mamw2011}
\begin{equation}
\label{eq:tprec}
  \tpr = \frac{M_{\bullet}}{\mcusp} T_{\rm orb} \, f(e)\,, 
\end{equation}
where $T_{\rm orb}$ is the orbital period of a star with semi-major
axis $a$, $\mcusp$ is the total stellar mass enclosed within
the orbit and
\begin{equation}
  f(e) = \frac{1 +\sqrt{1  - e^2}}{\sqrt{1-e^2}}\,.\nonumber
\end{equation}
$\tpr$ is of about $10^5\yr$ for the stars in the ring.  
For $\gamma = 3/2$, $\tpr$ is independent of $a$.  
For our adopted broken power-law
profile, $\tpr$ is only weakly dependent on distance (see
Fig.~\ref{fig:times}).

\subsection{Initial conditions}
 We adopt the outcomes of run~E by M12 as initial conditions for the
 simulations presented in this paper. In particular, M12 performed
 N-body/SPH simulations of a gas cloud evolving in the potential of a
 $M_{\bullet} = 3.5\times10^6\msun$ SMBH and a stellar cusp as
 described in Eq.~\ref{eq:cusp}.
\begin{figure}
  \begin{center}
    \includegraphics[width=8cm]{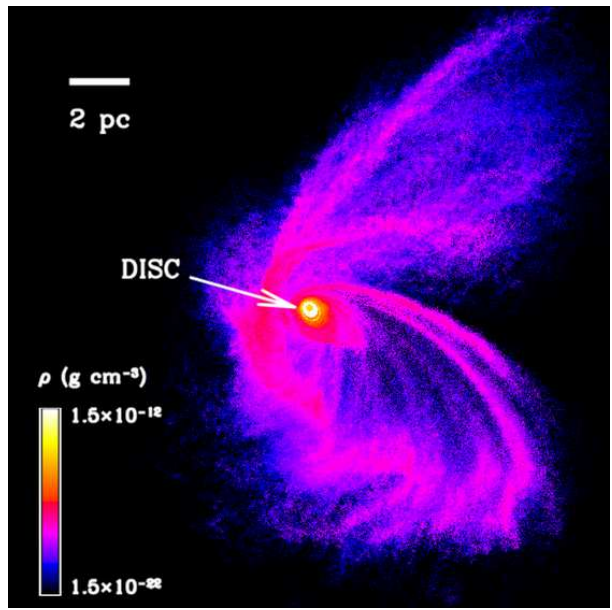}
  \end{center}
  \caption{Density map of the gas in run~E of M12 at
    $t=4.8\times 10^5\yr$. The simulation was projected in the plane
    where the gaseous disc (at the centre) is seen face-on. The box
    measures $20\pc$ per edge.}
  \label{fig:cloud}
\end{figure}
The impact parameter of the cloud with respect to the SMBH is
$10^{-2}\pc$ and the initial velocity is close to the escape velocity
from the SMBH at the initial distance ($25\pc$). The cloud is quite
massive ($1.3\times 10^5\msun$), initially spherical (with a radius of
$15\pc$), marginally self-bound and seeded with supersonic turbulent
velocities. The thermodynamic treatment in run~E includes radiative
cooling with accurate recipes for opacity \citep{boley2009} and an
irradiation temperature $T_{\rm irr}=100$ K, mimicking the background
temperature of the GC. Due to the SMBH shear, the molecular cloud is
disrupted in $\approx 10^5\yr$ and, because of its low orbital angular
momentum, part of the gas settles in a very small and dense disc
surrounding the SMBH (see Fig.~\ref{fig:cloud}). A fraction of the gas
in the disc fragments forming self-bound clumps (with density
$\ge{}2\times 10^{12}$ cm$^{-3}$). As a combined effect of the initial
cloud mass and of the floor temperature, the mass function of the
clumps is top-heavy, ranging between 1 and 60 M$_\odot{}$ after
$4.8\times10^5\yr$.

\begin{figure}
  \begin{center}
    \includegraphics[width=8cm]{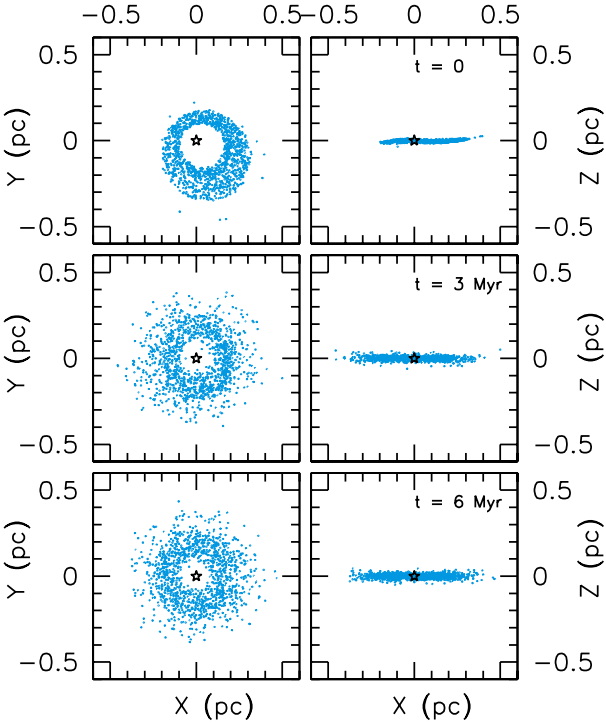}
  \end{center}
  \caption{Snapshots of the ring stars at different times during
      the evolution (from top to bottom $t=0,3,6\myr$, in a reference
      frame in which the stars lie in the $x-y$ plane. The star symbol
      marks the position of the SMBH.}
  \label{fig:snap}
\end{figure}

Here we assume that each self-bound clump formed in run~E of M12
becomes a star, without accreting any further gas particles after
$t=4.8\times 10^5\yr$\footnote{This assumption is quite conservative
  for the masses of stars, as it is realistic to expect further gas
  accretion in run~E of M12.}. Thus, we replace each gas clump with a
single star particle having mass equal to the total mass of the clump
itself, and position and velocity corresponding to those of the
centre-of-mass of the clump. The resulting stellar ring, with a total
mass of $\sim 4.3\times10^3\msun$, is shown in the top panels of
Fig.~\ref{fig:snap}. The edge-on view shows a small degree of warping.

In this paper, we cannot integrate the star particles together with
the gas particles, as it would be prohibitive to follow the gas for
$\gtrsim 1\myr$. Thus, we substitute the remaining gas particles with
an analytical potential (Eq.~\ref{eq:disc}) very close to the final
distribution of gas in run~E of M12 (see their figure 10). The total
mass of the gas disc within $0.4\pc$ (i.e. the outer edge of the
stellar ring) is $\sim10^4\msun$. For simplicity, the potential is
assumed constant in time. While this assumption may seem unrealistic,
it is justified by the fact that the evolution of the ring stars is
dominated by the SMBH and the stellar cusp, and the gas disc potential
represents a small correction to the total potential. We performed an
additional simulation excluding the gas disc potential and we found a
very similar evolution for the ring stars.

The SMBH is treated as a point-mass $N$-body particle, similarly to
the ring stars. However, given the small number of particles in the
simulations, the SMBH would be subject to an unrealistically large
motion at the center of the potential well \citep{merritt2005}, which
might affect the dynamics of the stars. For this reason, at each
integration step we force the SMBH to remain fixed at the center of
the reference frame.

\section{Results}
\label{sec:rotating}
We let the ring of stars evolve for $10\myr$ under the effect of the
gravitational potential generated by the SMBH, the stellar cusp, and
the gas disc. Due to the presence of extended mass (stars and gas)
within the orbit of each ring star, stars precess in their orbital
plane. Because of its much higher mass, the stellar cusp is
responsible for most of the precession.  While the dependence of the
precession time on distance is rather weak (see Fig.~\ref{fig:times}),
the dependence on eccentricity may be significant.  This implies a
different precession time for stars with different eccentricity, and
it results in a quick loss of coherence in the ring, as can be seen
from the snapshots at $3\myr$ and $6\myr$ shown in
Fig.~\ref{fig:snap}. Only in the absence of a stellar cusp would the
ring evolve coherently.

\begin{figure}
  \begin{center}
    \includegraphics[height=8cm, angle=-90]{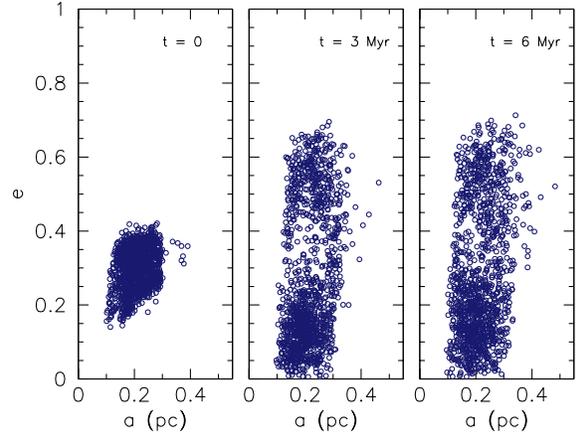}
  \end{center}
  \caption{Orbital elements of the ring stars at the start of the
    integration (left panel), after $3\myr$ (middle panel) and
      after $6\myr$ of evolution (right panel).}
  \label{fig:orb}
\end{figure}

\begin{figure}
  \begin{center}
    \includegraphics[height=8cm, angle=-90]{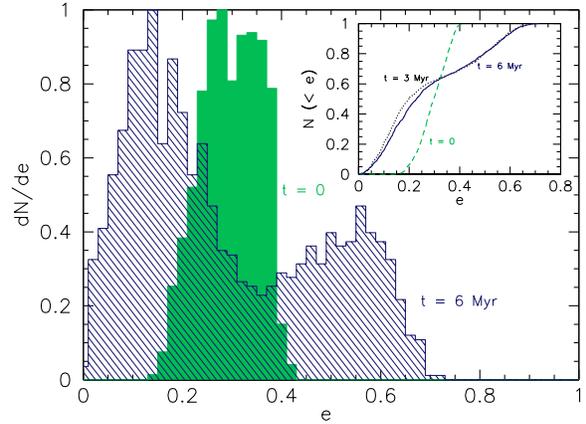}
  \end{center}
  \caption{Eccentricity distribution of the ring stars at the start of
    the integration (filled area) and after $6\myr$ of evolution
    (hatched area).  The insert in the top-right corner shows the
    cumulative distribution of eccentricities at different times: at
    the beginning of the integration (dashed line), after $3\myr$
    (dotted line) and after $6\myr$ (solid line).}
  \label{fig:ecc}
\end{figure}

The orbital elements of all stars are shown in Fig.~\ref{fig:orb}, at
different times during the integration.
While the semi-major axes distribution remains unchanged
during the evolution, the distribution of eccentricities evolves toward
a bimodal distribution, as can be seen in Fig.~\ref{fig:ecc}.
While the initial distribution of eccentricities shows a single peak
at $e\sim 0.3$, the distribution after $6\myr$ has a primary peak at
$e\sim 0.1$ and a secondary peak at $e\sim 0.5$, with a tail that
extends to about $e\sim0.7$. The cumulative distributions of
eccentricities are shown in the insert.

\begin{figure}
  \begin{center}
    \includegraphics[height=8cm, angle=-90]{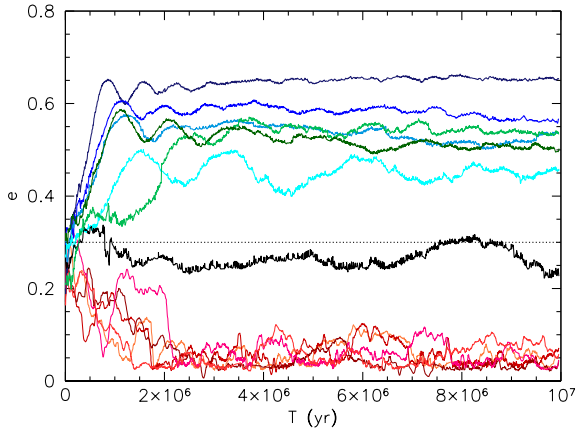}
  \end{center}
  \caption{Eccentric orbit instability in a random subset of
    stars. The timescale for the process is about $1\myr$.}
  \label{fig:inst}
\end{figure}

We attribute this behaviour to the eccentric disc instability which
was predicted to occur in eccentric discs subject to cusp precession
\citep{madigan2009}. Due to the fact that $f(e)$ in Eq.~\ref{eq:tprec}
is a monotonically increasing function of eccentricity, an orbit with
initial eccentricity in excess of the average will have a longer
precession time, which will result in a stronger torque from the other
stars in the ring. The torque will reduce its angular momentum and
therefore cause an increase in eccentricity. The opposite effect will
happen to stars with orbits initially more circular than the average. The
instability is illustrated in Fig.~\ref{fig:inst}, which shows the
time evolution of the orbital eccentricity for a random subset of
stars in the ring.
While stars initially more eccentric than the average ($\sim 0.3$)
tend to become more eccentric, stars initially less eccentric than
the average tend to circularise. The observed timescale for the
process is of only $\sim 1\myr$. This is in agreement with the
predicted timescale \citep{madigan2009}
\begin{equation}
\label{eq:tecc}
\tecc \sim \tpr \left(\frac{\mcusp}{\mring}\right)^{1/2} \left[
  e \sqrt{1-e^2} \frac{d}{de} \left(\frac{1}{f(e)}\right)\right]^{-1/2}\,,
\end{equation}
where $\mcusp$ and $\mring$ are the cusp mass and the ring mass,
respectively.  In our case, $30 \lesssim \mcusp / \mring \lesssim
  60$ for $ 0.1 \le r/\pc \le 0.4$ and $\tecc \sim 5-10$ for a
  circular orbit.  As the
ring loses coherence after this time, torques become weaker and
changes in the eccentricities become less pronounced.

This distribution is consistent with, though not identical to, the one
found for the CW disc. \citet{bartko2009} perform a statistical
analysis of the candidate disc members, assuming Keplerian orbits in
the gravitational potential of the SMBH and the stellar cusp, and
derive distributions for the orbital elements of the stars.  While
these distributions are subject to many uncertainties, the main
properties of the eccentricity distribution appear robust.  In
particular, they report a bimodal distribution for the reconstructed
eccentricities with peaks at $e \sim 0.3$ and $0.9$, though they
caution that the high eccentricity stars may not dynamically belong to
the disc.  Once corrected for projection effects, the mean
eccentricity in the CW disc is $\sim 0.37$, only slightly higher than
what we find.  The difference is likely due to the eccentricity with
which the stars form from the gas disc, which in turn depends on the
orbital parameters of the cloud. An average eccentricity of $\gtrsim
0.4$ could be produced starting with an even more radial orbit for the
molecular cloud, which might result from cloud-cloud collisions near
the GC (e.g. \citealt{hobbsnayakshin2009}).  Very large eccentricities
($\gtrsim 0.99$) like those found by \citet{madigan2009}, however,
seem hard to produce even starting from a more eccentric gas disc.
This has important implications for models of the origin of the
S-cluster in the GC.  For tidal disruption and capture to occur, a
binary must come at least as close to the SMBH as the tidal radius.
\citet{madigan2009} find that for a $4\times10^3\msun$ disc with
initial eccentricity of $0.6$ the maximum eccentricity attained by the
stars is about 0.9, and yet there are no stars with a pericentre
distance small enough to be tidally disrupted. Only discs more massive
than $10^4\msun$ and with initial eccentricities in excess of 0.6
result in a significant number of tidal disruptions.  We conclude that
the S-stars cannot have formed from a disc like the one produced here.

\begin{figure}
  \begin{center}
    \includegraphics[height=8cm, angle=-90]{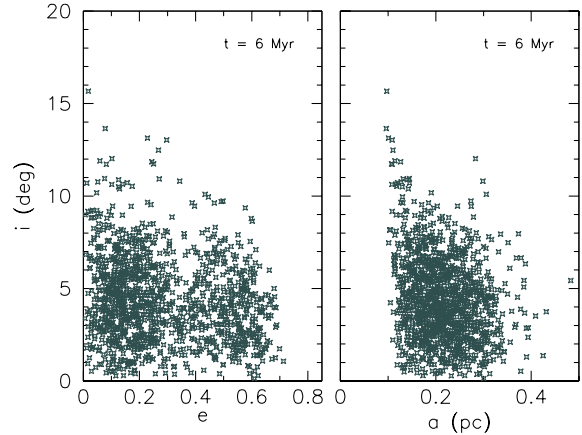}
  \end{center}
  \caption{Inclinations of the ring stars after $6\myr$ of evolution
    as a function of eccentricity (left panel) and semi-major axis
    (right panel). The inclination is calculated as the angle between
    the angular momentum of the star and the z-axis, which is parallel
    to the initial total angular momentum of the ring.}
  \label{fig:incl}
\end{figure}
In addition to finding smaller eccentricities than in some of
Madigan's simulations, we find smaller inclinations with respect to
the initial ring plane. Inclinations after $6\myr$ are generally
smaller than $\sim 10$ degrees and the maximum we record is $\sim 16$
degrees. Inclinations do not seem to correlate with eccentricities
(see Fig.~\ref{fig:incl}), though we note that if we divide the sample
of stars in two groups, having $e < 0.3$ and $e \geq 0.3$, the
inclinations in the two groups evolve in a different way. A 2D
Kolmogorov-Smirnov test gives probability values of 0.07 at the start
of the simulation, 0.01 after $3\myr$ and $2.1\times10^{-6}$ after
$6\myr$ that the inclinations in the two groups originate from the
same distribution. This difference is likely due to differential
heating between more eccentric and less eccentric stars. On the other
hand, Fig.~\ref{fig:incl} shows a trend for higher inclinations to
develop for stars closer to the SMBH. This is a result of the initial
warping present in the stellar ring, which in turn reflects the
warping of the gas disc. Larger inclinations might originate from more
strongly warped gas discs, but additional SPH simulations would be
required to test this scenario. The rms inclination grows slowly with
time, roughly as $t^{1/8}$, while the rms eccentricity remains
approximately constant at $e \approx 0.3$.  This scaling is
  somewhat flatter than what is found in \citet{cuadra2008} and
  \citet{hs2012}, and it may be attributed to the different
  initial conditions adopted in this work, in particular to the non
  zero mean initial eccentricity and/or warping of the stellar ring.
The modest inclinations achieved by the simulated stars imply that the
single cloud infall scenario cannot account for the high inclination
of some of the young stars observed in the GC (those claimed to belong
to a counterclockwise disc).  A cloud-cloud collision has been shown
to produce significantly warped stellar discs and additional
filamentary structures at large inclinations with respect to the disc
\citep{hobbsnayakshin2009}. Alternatively, the interaction between two
approximately coeval and highly inclined stellar discs may provide an
explanation for the existence of stars at high inclination
\citep{lb2009}.

Finally, we find a very modest heating of the stellar ring during the
evolution. This is likely due to the absence of heating from the
stellar cusp in our simulations, which would result in an increase in
the mean ring height \citep[see e.g.][]{pgma2008}.

\section{Discussion and conclusions}
\label{sec:concl}

We studied the evolution of the ring of stars formed in the GC from
fragmentation of the gas disc deposited by an inspiralling molecular
cloud. Our calculation is the first of its kind to follow, in a
consistent manner, both the star formation phase in the gaseous disc
and the long term evolution of the stellar disc. We find that, while
the ring retains the original distribution of semi-major axes, and
therefore also the initial inner and outer radius, the distribution of
eccentricities evolves in time due to the onset of the eccentric disc
instability. Torques exerted by other stars in the ring result in a
change in the magnitude of the angular momentum and, as a consequence,
in the eccentricity. As stars evolve away from the average
eccentricity, a bimodal distribution is established, with a primary
peak at $e\sim 0.1$, a secondary peak at $e \sim 0.5$, and a tail that
extends to $e \sim 0.7$. This is qualitatively consistent with the
distribution found for the CW disc stars.  We predict that a
quantitative agreement would require a molecular cloud with an initial
orbit more radial than in run~E of M12, but a new suite of
hydrodynamical simulations is necessary to test this prediction.  On
the other hand, even a better description of the cloud thermodynamics
and of the gas shocks taking place during the infall of the cloud
might affect the formation and thus the orbital properties of the
young stellar ring (see e.g. the discussion in \citealt{jg2011}, and
references therein).

Our simulations support the scenario in which the thin CW disc
originated from fragmentation of a moderately eccentric gas disc
resulting from the tidal disruption of a turbulent molecular cloud on
a low angular momentum orbit.  The origin of the WR/O stars at large
inclinations, as well as those belonging to the S-cluster, however,
seems to require additional physical processes, which are currently
under investigation.  In the case of the S-stars, relaxation processes
against the background spherical cusp of stars and remnants are unable
to explain the observed distribution of eccentricities within the
lifetime of the stars \citep{pgkma2009} and different mechanisms are
required \citep[for a review see e.g.][]{pg2010}.  Similarly,
\citet{cuadra2008} have shown that relaxation in the disc does not
produce eccentricities and inclinations as large as those observed for
the WR/O stars outside the CW disc.  While some outliers may be the
result of vector resonant relaxation between the disc and the stellar
cusp \citep{KT2011}, several authors have investigated the possibility
of an additional perturbing potential, which would produce Kozai-Lidov
type oscillations \citep{kozai1962,lidov1962} in the eccentricity and
inclination of the stars.   In particular, \citet{ssk2009}
  suggest that precession due to the gravity of the circumnuclear disc
  (CND) at the edge of the SMBH's sphere of influence might have
  driven some of the stars away from the CW disc.  An axisymmetric
  perturbation due the presence of an additional gas/stellar disc, an
  intermediate-mass black hole or the CND might have observable
  effects also on the stars in the CW disc.  \citet{lbk2008} studied
  the effect of a second highly inclined stellar disc, while
  \citet{ssk2009} and \citet{hsk2011} considered the effect of the
  CND.  \citet{lbk2008} found that, in the absence of a stellar
  cluster, precession and Kozai oscillations due to the presence of a
  second, inclined disc result in stars achieving high eccentricities.
  The inclusion of a stellar cusp, however, has been shown to damp
  Kozai oscillations in the disc \citep[e.g.][]{chang2009,lbk2009}, at
  least in the case of an analytic spherical cusp.  In this case, the
  overall influence of the cusp can be characterised by a decrease in
  the amplitude of the eccentricity and inclination oscillations, and
  by a shortening of their period
  \citep{KS2007,ssk2009}. \citet{chang2009} examined the effectiveness
  of the Kozai mechanism induced by stellar discs in the Galactic
  centre and found that Kozai oscillations are suppressed if the
  period for apsidal precession induced by the cusp is shorter than
  the period for Kozai oscillations. In the case of the two disc
  scenario, the mass of the stellar cusp in the GC is over an order of
  magnitude larger than the critical mass required to suppress the
  Kozai mechanism.

  In general, the damping of the Kozai mechanism
  depends on the parameters of the perturbing potential and on the
  orbital parameters of the stars, and oscillations might be important for
  certain sets of parameters. 
  The period of Kozai oscillations can be written as \citep{ssk2009}
  \begin{equation}
    \label{eq:tkozai}
  T_{\rm K}  = \frac{M_{\bullet}}{M_{\rm P}} \frac{R_{\rm
      P}^3}{a\sqrt{GM_{\bullet} a}} \sim \frac{M_{\bullet}}{M_{\rm P}}
  \left(\frac{R_{\rm P}}{a}\right)^3 T_{\rm orb}
  \end{equation}
  where $M_{\rm P}$ and $R_{\rm P}$ are the mass and radius of the
  perturbing axisymmetric potential.  If the perturber is a second
  disc at approximately the same distance as the CW disc from the
  SMBH, $R_{\rm P} \sim a$ and $T_{\rm K} / \tpr \approx \mcusp/M_{\rm
    P}$.
  If the perturber is the CND, then $R_{\rm P} \sim (5-20) a$.  In
  both cases, it seems unlikely that Kozai oscillations might dominate
  the evolution of the CW disc in the presence of a spherical cusp.
  The timescale for Kozai oscillations is also typically longer than
  the timescale for the eccentric disc instability (see
  Eq.\,\ref{eq:tecc}), suggesting that Kozai oscillations should not
  significantly affect the development of the eccentricity growth
  studied here, even if this is theoretically possible for certain
  combinations of parameters.

  \citet{lbk2009} explored the effect of a granular $N$-body cusp as
  opposed to an analytic smooth cusp.  They found that both a smooth
  and a granular cusp consistent with the one derived for the GC
  \citep{scho2007} suppress Kozai cycles in the two disc
  scenario. However, in the case of an $N$-body cusp, two-body and
  resonant relaxation effects may temporarily drive the eccentricty of
  some stars to large values. This effect is a function of the cusp
  granularity, i.e. of the mean and maximum stellar masses, and is
  most efficient in the case of a cusp of stellar remnants. Similar
  results were obtained by \citet{hs2012}.

  It is not yet clear whether secular evolution can play an important
  role under realistic conditions, i.e. a realistic gravitational
  potential and granularity of the stellar cusp. Shape and parameters
  of the potential of an additional disc and/or the CND are not well
  determined, and the exact composition of the GC cusp is not well
  known.  Further investigation of these issues is deemed necessary
  and will be presented in an upcoming work.

\section*{Acknowledgments}
We thank the referee Ladislav {\v S}ubr for useful comments that helped
improve the manuscript. The simulations were performed on the GPU
enabled machines at the Max-Planck Institute for Astrophysics in
Garching, Germany. MM acknowledges financial support from INAF through
grant PRIN-2011-1.

\bibliographystyle{mn2e}
\bibliography{biblio}

\end{document}